\documentclass[preprint,showpacs,amsmath,amssymb,floatfix,pre]{revtex4}
\usepackage{graphicx}
\begin{document}
\title{Multicritical behavior of the two-dimensional transverse Ising metamagnet
in a longitudinal magnetic field} 

\author{Denise A. do Nascimento}

\author{Josefa T. Pacobahyba}
\affiliation{Departamento de F\'{\i}sica, Universidade Federal de Roraima, 
BR 174, Km 12. Bairro Monte Cristo. CEP: 69300-000, Boa Vista-RR, Brazil}

\author{Minos A. Neto}
\email{minos@pq.cnpq.br}
\author{Octavio R. Salmon}
\affiliation{Departamento de F\'{\i}sica, Universidade Federal do Amazonas, 3000, Japiim,
69077-000, Manaus-AM, Brazil}

\author{J. A. Plascak}
\affiliation{Universidade Federal da Para\'{\i}ba, Centro de Ci\^encias Exatas e da Natureza - 
Campus I, Departamento de F\'{\i}sica - CCEN Cidade Universit\'aria 58051-970 - Jo\~ao Pessoa-PB, Brazil}
\affiliation{ Department of Physics and Astronomy, University of Georgia, 30602 Athens GA, USA}

\date{\today}

\begin{abstract}
\textbf{ABSTRACT}

Magnetic phenomena of the superantiferromagnetic Ising model in both uniform 
longitudinal ($H$) and transverse ($\Omega $) magnetic fields are studied by
employing a mean-field variational approach based on Peierls-Bogoliubov
inequality for the free energy. A single-spin cluster is used to get the
approximate thermodynamic properties of the model.
The phase diagrams in the magnetic fields and temperature ($T$) planes, namely,
$H-T$ and $\Omega-T$, are analyzed on an anisotropic square lattice for 
some values of the ratio $\alpha=J_{y}/J_{x}$, where $J_x$ and $J_y$ are the
exchange interactions along the $x$ and $y$ directions, respectively. Depending on 
the range of the Hamiltonian parameters, one has only second-order transition lines, 
only first-order transition lines, or first- and second-order transition
lines with the presence of tricritical points. The corresponding phase diagrams
show no reentrant behavior along the first-order transition lines at low
temperatures. These results are different from those obtained by using
Effective Field Theory with the same cluster size.

\end{abstract}

\maketitle

\section{Introduction\protect\nolinebreak}

Theoretical metamagnetic models are systems that have both antiferromagnetic and
ferromagnetic coupling interactions. 
At zero external field \cite{meijer1978,kincaid1975}, they undergo a 
second-order phase transition at the N\'eel temperature $T_N$. Besides the
usual nearest-neighbor staggered arrangements of the spins, the ordered phase
can be, for instance, ferromagnetic planes ordered antiferromagnetically to each
other, or even ferromagnetic chains ordered antiferromagnetically, where the
latter phase is usually called a superantiferromagnetic phase. At non-zero
field, applyied longitudinally to the direction of the magnetization, the
N\'eel transition temperature 
decreases as the magnitude of the external field increases and the line of
second-order 
transition comes  to an end at a finite temperature. Beyond this point, the transition 
is first order and ends up, at zero temperature, at a finite field
$H_c$. This phenomenon has been previously well described by Landau \cite{landau1937} and 
by Griffiths \cite{griffits1970}, where one has the presence of a tricritical point 
(the point joining first- and second-order transition lines).

On the other hand, through the years, experimental realizations of magnetic
materials exhibiting the above characteristics have been studied. 
For example, Chernyi and co-workers \cite{chernyi} have considered the kinetics 
of a magnetization process in quasi-one-dimensional Ising superantiferromagnet 
named as trimethylammonium 
cobalt chloride [(CH$_{3}$)$_{3}$NH]CoCl$_{3}\cdot$2H$_{2}$O, denoted by CoTAC,
belonging to a wider series of organometallic compounds with general chemical
formula [(CH$_{3}$)$_{3}$MX$_{3}$]CoCl$_{3}\cdot$2H$_{2}$O with M=Mn, Co, Ni,
Fe, and X=Br or Cl. However, one of the first studies known in the literature
on 
the effects of a longitudinal magnetic field in  superantiferromagnetic systems
has been done on the (C$_{2}$H$_{5}$NH$_{3}$)$_{2}$CuCl$_{4}$ compound
\cite{dejongh}.

It is worth mentioning that superfluid mixtures of the two helium isotopes 
$^{3}${He} and $^{4}${He}, and a 
class of anisotropic metamagnets such as FeCl$_{2}$, FeBr$_{2}$ and 
Ni(NO$_{3}$)$_{2}\cdot2$H$_{2}$O, although microscopically different at first
sight, show in
their thermodynamic behavior a striking similarity in that they exhibit a phase
diagram in which a line of $\lambda-$type transition points (second-order transition line) 
ends up at a
tricritical point (see,  for instance, reference \cite{kincaid1975}). 

Based on these real experimentations, we will treat herein a quantum version of a metamagnet 
conveying not only the basic features of the relevant classical interactions
above discussed, which leads to a tricritical behavior 
similar to the well known classical Blume-Capel \cite{bc} and BEG models
\cite{beg}, but also the inclusion of quantum fluctuations due to a transverse field, 
which will be in fact
important in the low temperature regime. Moreover, there has been, in recent
years, an extensive literature on competitive interaction models that present a
superantiferromagnetic (SAF) ordering, where, for instance, ferromagnetic
chains are coupled in an antiferromagnetic way in two dimensions.
According to these lines, the system to be treated herein corresponds to a 
two-dimensional spin-$1/2$ Ising
model with different exchange interactions along the two lattice directions, in
the presence of transverse and longitudinal magnetic fields. Additional
motivations to
study the Ising model is because it can be used to describe the
critical behavior of a broader class of materials, including easy-axis magnets,
binary alloys, simple liquids and their mixtures, polymer solutions, subnuclear
matter, etc. \cite{ising1,ising2,ising3}.

In the present case, the corresponding model can be described by the following
Hamiltonian
\begin{equation}
\mathcal{H} =
-J_{x}\sum_{{\left<i,j\right>}_x}\sigma
_{i}^{z}\sigma_{j}^{z}
+J_{y}\sum_{{\left<i,j\right>}_y}\sigma
_{i}^{z}\sigma_{j}^{z}
-H\sum_{i=1}^N\sigma _{i}^{z}-\Omega \sum_{i=1}^N\sigma_{i}^{x},
\label{1}
\end{equation}
where $\sigma_{i}^{\mu}$ is the $\mu(=x,y)$ Pauli spin-$1/2$ operator component
at site $i$ on a square lattice of $N$ sites, $J_{x}>0(J_{y}>0)$ is the exchange
coupling along the $x(y)$ axis, the first and second sum are over
nearest-neighbors along the $x$ and $y$ axis, respectively, $H$ is the 
longitudinal magnetic field, and $\Omega$ is the transverse magnetic field. 
The corresponding ordered state is composed by a superantiferromagnetic 
phase, and is characterized by a parallel spin orientation in the
$x$ direction, and an antiparallel spin orientation in the $y$
direction, therefore exhibiting a kind of N\'eel order between two sublattices
of linear chains that can be denoted by $A$ and $B$.

The classical version of model (\ref{1}), i.e. $\Omega=0$, on anisotropic square
lattices has been
investigated by using a modified mean-field theory, in which the intrachain
interaction is treated exactly and the
interactions between chains are taken into account in a mean-field way (linear
chain approximation - LCA)
\cite{stout,hone,sato,lca,pla2}. Several other approaches, such as usual mean-field
approximation (MFA) \cite{garrett,ziman,rottman1990}, effective-field theory
(EFT) \cite{minos2006,zukovic,sousa2013}, mean-field
renormalization group (MFRG) \cite{slotte}, effective-field renormalization
group (EFRG) \cite{neto2004}, Monte Carlo simulations (MC)
\cite{viana2009,landau,landau1976,ferrenberg,otavio}, and high-temperature
series expansion (SE) \cite{bienenstock} have also been applied to this
classical model. Although these approaches agree with the overall phase
diagram in two dimensions, LCA \cite{lca} and EFT \cite{minos2006}
approaches produce a reentrant behavior in the first-order transition line. The
same reentrant behavior is obtained for the three-dimensional model by the EFT
with four spins \cite{sousa2013}. 

On the other hand, the quantum version of the model, given by the Hamiltonian
(\ref{1}), has not been ubiquitously treated in the literature as its classical counterpart.
Nevertheless, phase diagrams and some thermodynamic properties have been
obtained by using the EFT \cite{deni1,deni2}, and a pair approximation for the
free energy (the latter case only for $J_y<0$ \cite{pla}). It has been noted
that only second-order phase transitions are present in the computed phase
diagrams.

In the present paper, using the MFA based on a variational method for the free-
energy, we investigate the phase diagram behavior of the spin-$1/2$ Ising
superantiferromagnet in the presence of both longitudinal and transverse
external fields. We would like to seek out not only the effect of the quantum
fluctuations in the model but also the presence or not of the tricritical
points and first-order transitions, since the EFT has only given a second-order
character. 

The remaining of the paper is organized as follows. In the next section we
outline the formalism and its application to the transverse Ising
superantiferromagnet in the presence of a longitudinal magnetic field; in Sec.
III we discuss the results and present some final comments.

\section{Formalism}

We will treat model (\ref{1}) by the mean-field approximation (MFA) using a
variational method based on Peierls-Bogoliubov inequality, which can be
formally written as
\begin{equation}
F\left(\mathcal{H}\right)\leq F_{0}\left(\mathcal{H}_{0}\right)+\left\langle
\mathcal{H}-\mathcal{H}_{0}\right\rangle_0\equiv\Phi(\gamma), 
\label{2}
\end{equation}
where $F$ and $F_{0}$ are free energies associated with two systems defined by
the Hamiltonians $\mathcal{H}$ and $\mathcal{H}_{0}(\gamma)$, 
respectively, the thermal average $<...>_0$ should be taken in relation to the canonical
distribution associated with the trial Hamiltonian $\mathcal{H}_{0}(\gamma)$, with $\gamma$ 
standing for the variational parameters. The approximated free energy $F$ is
then given by the minimum of $\Phi(\gamma)$ with respect to $\gamma$, i.e.
$F\equiv\Phi_{min}(\gamma)$.

The trial Hamiltonian $\mathcal{H}_{0}$ is chosen as free spins, distributed in
two different sublattices $A$ and $B$. Each sublattice consists of
ferromagnetic linear chains coupled antiferromagnetically with two neighboring
chains. We then have
\begin{equation}
\mathcal{H}_0 =
-\gamma_A\sum_{i\subset A}\sigma _{i}^{z}-\Omega \sum_{i\subset A}\sigma_{i}^{x}
-\gamma_B\sum_{i\subset B}\sigma _{i}^{z}-\Omega \sum_{i\subset B}\sigma_{i}^{x},
\label{h0}
\end{equation}
where $\gamma_A$ and $\gamma_B$ are two variational parameters.

It is not difficult to compute the right hand side of Eq. (\ref{2}) and,
after minimizing $\Phi(\gamma)$, the variational parameters $\gamma_A$ and
$\gamma_B$ can be written as a function of the sublattice magnetizations $m_{A}$
and $m_{B}$. The approximated
mean-field Helmholtz free energy per spin, $f=\frac{\Phi}{N}$, can thus be
written as
\begin{eqnarray}
f &=&\frac{t}{2}\ln\left\lbrace 4\cosh\frac{1}{t}\sqrt{\left(h+2m_{A}-2\alpha 
m_{B}\right)^{2}+\delta^{2}}\cosh\frac{1}{t}\sqrt{\left(h+2m_{B}-2\alpha
m_{A}\right)^{2}+\delta^{2}}\right\rbrace 
\label{fe}
\\
&&-\alpha m_{A}m_{B}+\frac{1}{2}\left(m_{A}^{2}+m_{B}^{2}\right), 
\label{3}
\end{eqnarray}%
with the corresponding sublattice magnetizations $m_{A}$ and $m_{B}$ given by
\begin{equation}
m_{A}=\frac{h+2\left(m_{A}-\alpha m_{B}\right)}{\sqrt{\left(h+2m_{A}-2\alpha
m_{B}\right)^{2}+\delta^{2}}}\tanh\frac{1}{t}\sqrt{\left(h+2m_{A}-2\alpha
m_{B}\right)^{2}+\delta^{2}},
\label{4}
\end{equation}%
and
\begin{equation}
m_{B}=\frac{h+2\left(m_{B}-\alpha m_{A}\right)}{\sqrt{\left(h+2m_{B}-2\alpha
m_{A}\right)^{2}+\delta^{2}}}\tanh\frac{1}{t}\sqrt{\left(h+2m_{B}-2\alpha
m_{A}\right)^{2}+\delta^{2}},
\label{5}
\end{equation}%
where $t=k_{B}T/J_{x}$, $h=H/J_{x}$, $\delta=\Omega/J_{x}$
and $\alpha=J_{y}/J_{x}$ is the ratio between ferromagnetic and
antiferromagnetic interactions. For a given value of the set of parameters $t$,
$h$, $\delta$ and $\alpha$, Eqs. (\ref{4}) and (\ref{5}) are numerically solved
for $m_{A}$ and $m_{B}$ and, when several solutions are found, the stable phase
will be the one that minimizes the free energy $f$.
In this context, the equilibrium state corresponds always to the minimum value of
$f$ with respect to $m_{A}$ and $m_{B}$. 

For a metamagnetic system, it is more convenient to 
formulate the problem in terms of the total $m$ and the staggered $m_s$
magnetizations, which are defined as $m=(m_{A}+m_{B})/2$ and
$m_{s}=(m_{A}-m_{B})/2$. Therefore, we can rewrite eqs. 
(\ref{3}), (\ref{4}) and (\ref{5}) as
\begin{eqnarray}
f
&=&-\frac{t}{2}\ln4\sum_{p=0}^{1}\left\lbrace\cosh\frac{1}{t}\sqrt{\left[
h+2\left(1-\alpha\right)m-2\left(-1\right)^{p}\left(1+
\alpha\right)m_{s}\right]^{2}+\delta^{2}}\right\rbrace 
\nonumber
\\
&&+\left(1-\alpha\right)m^{2}+\left(1+\alpha\right)m_{s}, 
\label{6}
\end{eqnarray}%
\begin{eqnarray}
m &=&\frac{1}{2}\sum_{p=0}^{1}\frac{h+2m\left(1-\alpha\right)-2m_{s}\left(1+
\alpha\right)}{\sqrt{\left[h+2m\left(1-\alpha\right)-2\left(-1\right)^{p}
\left(1+\alpha\right)m_{s}\right]^{2}+\delta^{2}}}
\nonumber
\\
&&\tanh\frac{1}{t}\sqrt{\left[h+2m\left(1-\alpha\right)-2\left(-1\right)^{p}
\left(1+ \alpha\right)m_{s}\right]^{2}+\delta^{2}}, 
\label{7}
\end{eqnarray}%
\begin{eqnarray}
m_{s} &=&\frac{1}{2}\sum_{p=0}^{1}\left(-1\right)^{p}\frac{h+2m\left(1-
\alpha\right)-2m_{s}\left(1+\alpha\right)}{\sqrt{\left[
h+2m\left(1-\alpha\right)-2\left(-1\right)^{p}\left(1+\alpha\right)m_{s}\right]^
{2}+\delta^{2}}}
\nonumber
\\
&&\tanh\frac{1}{t}\sqrt{\left[h+2m\left(1-\alpha\right)-2\left(-1\right)^{p}
\left(1+ \alpha\right)m_{s}\right]^{2}+\delta^{2}}, 
\label{8}
\end{eqnarray}%
from which one can obtain the frontiers separating
the SAF and P phases (the paramagnetic phase P consists of all spins aligned
with the external longitudinal field). Although the first-order transition line 
must be computed by numerically seeking the minimum of the free energy, the 
second-order transition line, as well as the location of the
tricritical point, can be obtained through a Landau expansion of the free
energy, given by Eq. (\ref{6}), in a power series of the order parameter
$m_{s}$. After a lengthy algebra the final result can be written as
\begin{equation}
f\left(t,h,\delta,\alpha;m_{s}\right)=\sum_{k=0}a_{2k}\left(t,h,\delta,
\alpha\right)m_{s}^{2k},
\label{9}
\end{equation}
from which we get the second-order transition lines when $a_2=0$ and $a_4>0$,
and the tricritical point when $a_{2}=0$, $a_{4}=0$ and $a_{6}>0$ (see, for
instance, Refs. \cite{gul,sousa2013}). In present case, $a_0$ is not important
for our purposes, so the $a_2$,
$a_4$, and $a_6$  coefficients  can be written as
\begin{equation}
 a_{2}=\frac{4\gamma_{2}^{2}}{\lambda}\left\lbrace \left[\frac{\beta\gamma_{1}^{2}}{\lambda}+1\right]-\frac{4\gamma_{1}^{2}}{\lambda}\left[\beta x + \frac{x}{\lambda}\right]\right\rbrace+2(1+\alpha),
\label{a2}
\end{equation}
\begin{eqnarray}
a_{4} &=&\frac{1}{\beta}\left(\frac{2\beta\gamma_{1}\gamma_{2}}{\lambda}\right)^{4}\left(x^{2}-1\right)-\frac{12\beta\gamma_{2}^{4}}{\lambda^{3}}\left(\lambda-x+\lambda x\right) + \frac{24\beta x^{2}\gamma_{2}^{4}}{\lambda^{6}}\left(\gamma_{1}^{4}-\lambda^{4}\right) \\ \nonumber
&& + \frac{48\beta^{2}\gamma_{1}^{2}\gamma_{2}^{4}}{\lambda^{3}}\left(x^{3}-x\right)-24\left(\frac{\beta\gamma_{1}\gamma_{2}}{\lambda}\right)^{4}\left[\frac{3}{\beta}\left(\frac{x}{\beta\gamma_{1}}\right)^{2} + \left(1+\frac{1}{\beta}\right)x^{4}-\left(1+\beta x^{2}\right)\frac{1}{\beta^{2}}\right]\\ \nonumber
&& +\frac{24}{\beta^{2}\lambda}\left(\frac{\beta\gamma_{1}\gamma_{2}}{\lambda}\right)^{4}\left[x^{3}-\frac{x}{(\gamma_{1}\gamma_{2})^{2}}-\beta x\right]+\frac{36}{\beta^{3}\gamma_{1}^{2}}\left(\frac{\beta\gamma_{1}\gamma_{2}}{\lambda}\right)^{4}\left[2\left(1-\frac{1}{\beta\lambda}\right)+\left(\frac{\gamma_{1}x}{\beta\lambda}\right)^{2}\right] \\ \nonumber 
&& -\frac{60\beta\gamma_{1}^{4}\gamma_{2}^{4}}{\lambda^{6}}\left(1-\frac{x}{\lambda\gamma_{2}^{2}}\right), 
\label{a4} 
\end{eqnarray}%
\begin{eqnarray}
a_{6} &=&(3780x)\frac{\gamma_{1}^{6}\gamma_{2}^{6}}{\lambda^{11}}-(1200x^{2})\frac{\gamma_{1}^{6}\gamma_{2}^{6}}{\lambda^{10}}+\Psi_{1}(\gamma_{1},\gamma_{2},\beta,x)\mathcal{O}\left(\frac{1}{\lambda^{9}}\right) \\ \nonumber
&& +\Psi_{2}(\gamma_{1},\gamma_{2},\beta,x)\mathcal{O}\left(\frac{1}{\lambda^{8}}\right)+\Psi_{3}(\gamma_{1},\gamma_{2},\beta,x)\mathcal{O}\left(\frac{1}{\lambda^{7}}\right)+\Psi_{4}(\gamma_{1},\gamma_{2},\beta,x)\mathcal{O}\left(\frac{1}{\lambda^{6}}\right) \\ \nonumber
&& +\Psi_{5}(\gamma_{1},\gamma_{2},\beta,x)\mathcal{O}\left(\frac{1}{\lambda^{5}}\right)+\Psi_{6}(\gamma_{1},\gamma_{2},\beta,x)\mathcal{O}\left(\frac{1}{\lambda^{4}}\right)+120(x+2x^{2})\frac{\beta^2\gamma_{2}^{6}}{\lambda^{3}},
\label{a6}
\end{eqnarray}%
where $\lambda^{2}=\gamma_{2}^{2}+\delta^{2}$, $\gamma_{1}=h+2(1-\alpha)m^{2}$, $\gamma_{2}=2(1+\alpha)$ and $x=\tanh(\beta\lambda)$. The functions $\Psi_{i}$, $i=2,~3,...,6$, are rather
lengthy to be reproduced here.

\section{Results and Discussion}

Depending on the value of the Hamiltonian parameters, the model (\ref{1}) has first- or 
second-order phase transitions from  the superantiferromagnetic phase (SAF) to the paramagnetic
phase (P). For $\Omega=0$, one has the well known mean-field results for the
classical model, in which the critical 
temperature decreases from $t_c=2+2\alpha$ as the longitudinal field increases and, at $T=0$ 
(ground-state), a first-order transition occurs at $h_{c}=2\alpha$. The tricritical point
is located at $t_t=2.667$ and $h_t=1.756$ for the isotropic lattice $\alpha=1$.
It would then be  quite interesting to see the global effects of the quantum
fluctuations, which are driven by the transverse field, on the corresponding
phase diagram.
\begin{figure}[htbp]
\centering
\includegraphics[width=7.0cm,height=7.0cm]{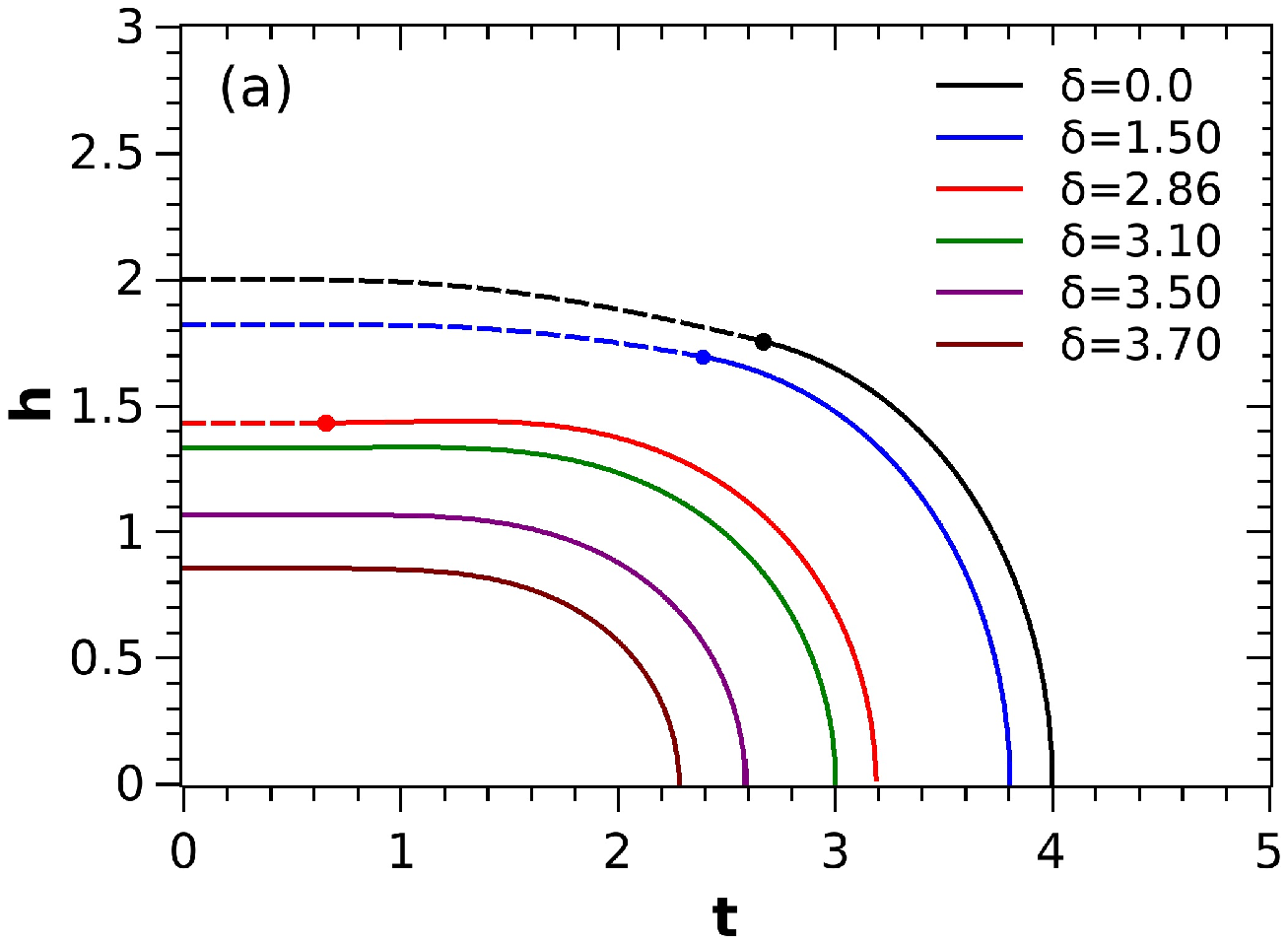}
\hspace{0.5cm}
\includegraphics[width=7.0cm,height=7.0cm]{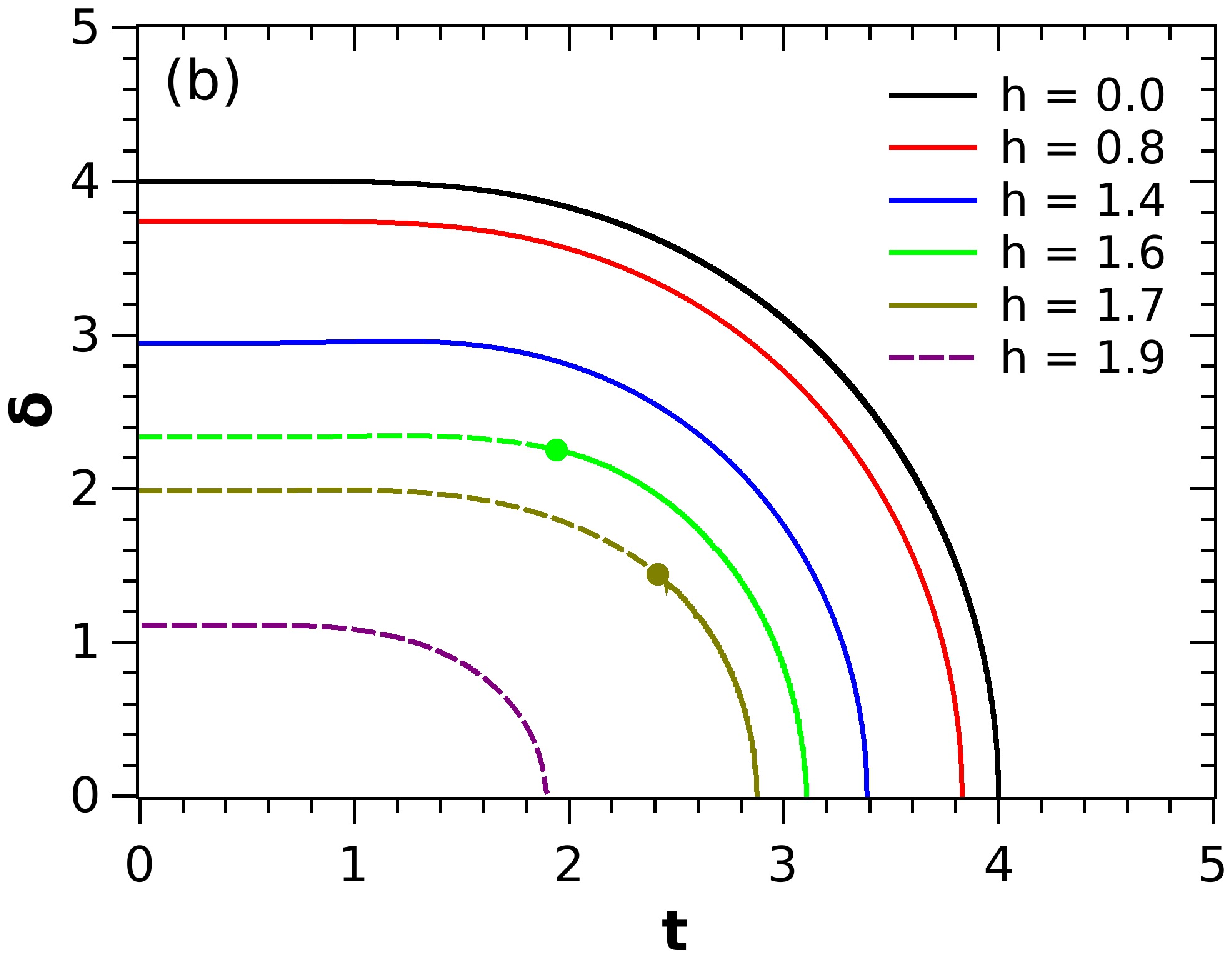}
\caption{Phase diagrams for the isotropic square lattice ($\alpha=1.0$). (a)
In the reduced temperature $t$ versus reduced longitudinal magnetic field $h$ plane, 
for several values of $\delta$. (b) In the reduced temperature $t$ versus the reduced 
transverse field $\delta$, for several values of $h$.} 
\label{mfa_diagrams_1}
\end{figure}

In Figs. \ref{mfa_diagrams_1}(a) and \ref{mfa_diagrams_1}(b) we have the phase
diagrams in the $h-t$ (for some selected
values of $\delta$) and $\delta-t$ (for some selected
values of $h$) planes,  in the isotropic lattice case $\alpha=1$. 
From Fig. \ref{mfa_diagrams_1}(a) we can see that as
the transverse field increases, the transition temperature decreases. This is a
result of the quantum fluctuations destroying not only the
superantiferromagnetic order among chains, but also the corresponding
ferromagnetic order inside each chain. The first-order
line and the tricritical point survive up to $\delta\simeq 2.86$ (this value,
as well as the ones given below, can be numerically obtained with much more
precision. However, for the sake of simplicity, we will only present them
with two decimal digits). For $2.87 \lesssim\delta\le4 $ the transition
is always second order, and quantum phase transitions start to develop at $T=0$, while for 
$\delta>4$ the system is always in the paramagnetic phase.

Fig. \ref{mfa_diagrams_1}(b) presents the corresponding phase diagram in the $\delta-t$ plane for
several values of $h$ and again $\alpha=1$. For $h=0$, we have the mean-field solution of 
the transverse Ising model, where the critical temperature goes to zero at the known value 
$\delta=4$. As expected, by increasing the longitudinal field the transition temperature
decreases. For $ 0<h\lesssim1.48$ the transition is always second order.
Tricritical points and 
first-order transition lines, at low temperatures, appear for
$1.49\lesssim h\lesssim1.80$ (in this region,
the critical quantum phase transition is suppressed). In the range
$1.81\lesssim h\le2$, only first-order
transition is observed and, for $h>2$, the system is always in the paramagnetic phase for any
value of the temperature.

\begin{figure}[htbp]
\centering
\includegraphics[width=7.0cm,height=7.0cm]{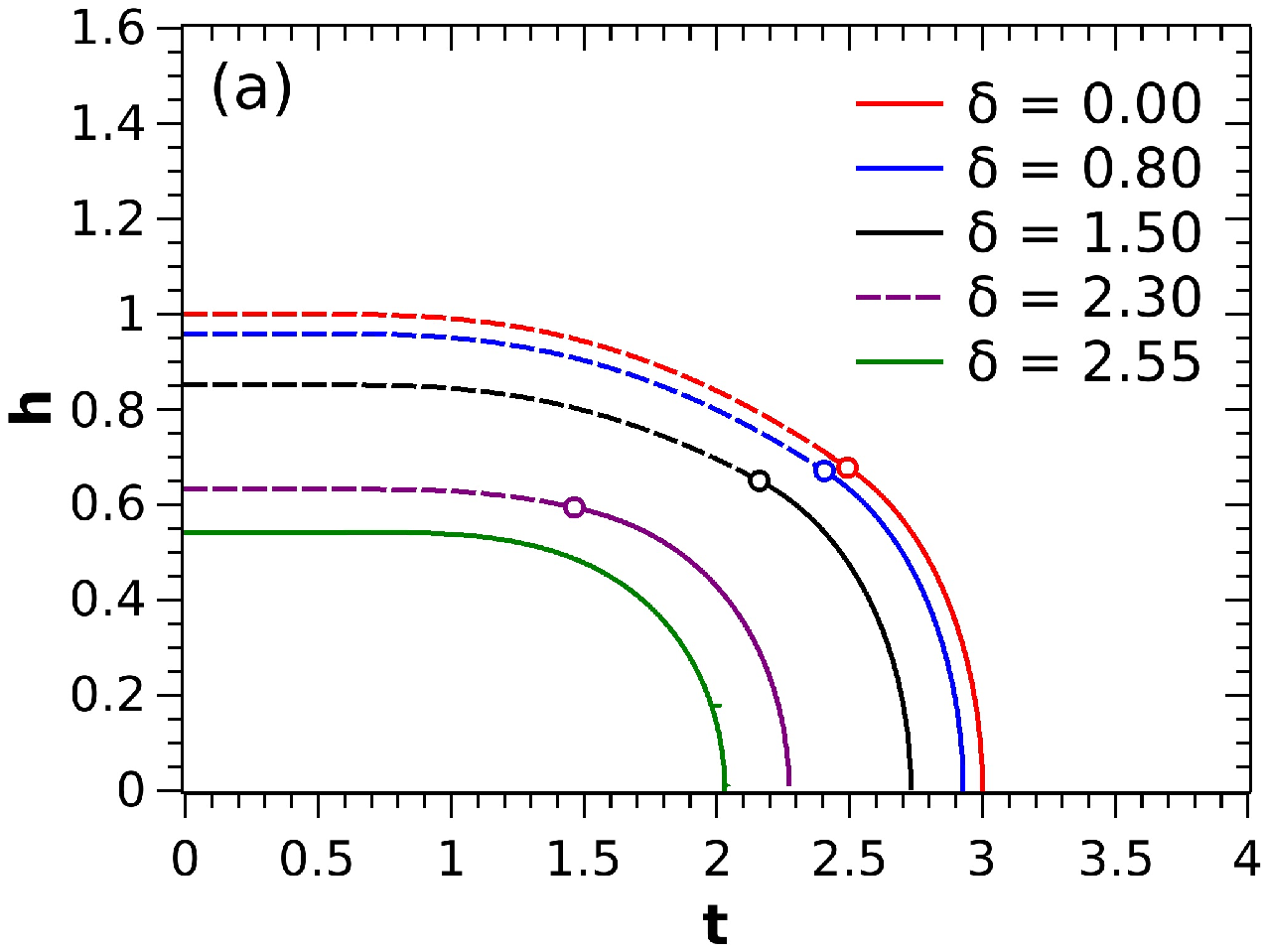}
\hspace{0.5cm}
\includegraphics[width=7.0cm,height=7.0cm]{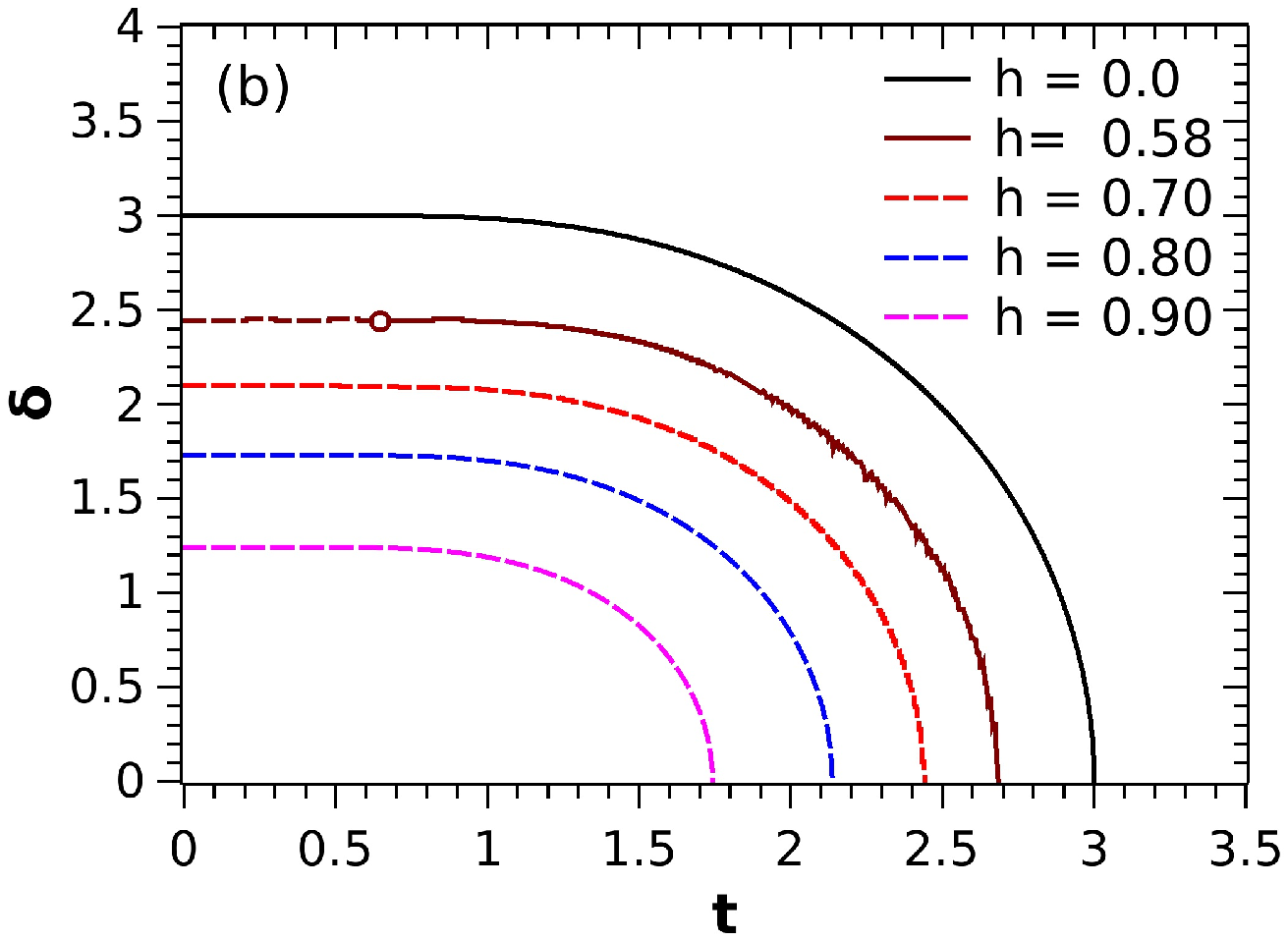}
\caption{The same as Fig \ref{mfa_diagrams_1} for $\alpha=0.5$.} 
\label{mfa_diagrams_2}
\end{figure}

In Figs. \ref{mfa_diagrams_2}(a) and \ref{mfa_diagrams_2}(b) we have the phase diagrams for
the anisotropic lattice $\alpha=0.5$. The general behavior is the same, we have only changes
in the values of the corresponding Hamiltonian parameters. For instance, in the $h-t$ plane
the tricritical point survives for $\delta\lesssim2.31$, for
$2.32\lesssim\delta\le3$ only second-order transition
lines are present and for $\delta>3$ the system is always in the paramagnetic phase. In the
$\delta-t$ plane one has always second-order transitions for $ 0<h\lesssim0.60$,
tricritical points and first-order transition lines appear for $0.61\lesssim
h\lesssim0.69$, only first-order transitions occur for $0.70\lesssim h\le2$, and
the paramagnetic phase is always stable for $h>2$. Note here the smaller range
of $h$ for having first-order and tricritical points in this plane phase
diagram.

The same trend is achieved for still smaller values of $\alpha$, where a
crossover from the two-dimensional behavior to the one-dimensional behavior is
observed when $\alpha\rightarrow0$. For this quasi-one-dimensional model,
however, the range of the parameters in order to observe multicritical
phenomena gets narrower, and it is sometimes difficult to numerically access it.
Moreover, as the present one-spin mean-field approach provides a final
transition temperature even in one dimension,
better procedures should be very welcome to treat the present model.
Nevertheless, from the present results, one can clearly see that the results are
indeed different from those obtained by employing EFT \cite{deni1,deni2} and no
reentrant behavior is observed in the first-order transition lines at low
temperatures. The same qualitative results are expected for the model in
three-dimensions, where the chains are arranged in a staggered way.

\section{Conclusions}

In summary, we investigated the anisotropic two-dimensional nearest-neighbor Ising model with competitive interations in an 
uniform longitudinal and traverse fields by using the MFA approach. We obtained the phase diagrams in the $h-T$ and $\delta-T$ 
planes varying the value of $\alpha$, where the critical frontier separates the SAF order with the paramagnetic disorder. 

At zero temperature, the critical field is exactly obtained, so  $h_{c}=2+2\alpha$. For a $\delta=0$, the model reproduces the 
classic result \cite{minos2006}. We see that there are the same trend is achieved for still smaller values of $\alpha$, where 
a crossover from the two-dimensional behavior to the one-dimensional behavior is observed when $\alpha\rightarrow0$. 

We showed also that for this approach appear in the phase diagrams lines of the first order as well as tri-critical points, and 
that the first order lines are derived from the construction of Maxwell. Furthermore, the investigations of this three-dimensional 
model are expected to show many characteristic phenomena as, for example, the reentrant behavior. This will be discussed in future 
works.

\textbf{ACKNOWLEDGEMENT}

This work was partially supported by FAPEAM and CNPq (Brazilian Research Agencies).

\end{document}